\documentclass[iop]{emulateapj}

\usepackage{natbib}
\usepackage{amsmath}

\newcommand{\gri}{\protect\hbox{$gri$} }

\newcommand{\jhks}{\protect\hbox{$J\!H\!K_{s}$} }

\newcommand{\about}{$\sim\!\!$~}
\newcommand{\kms}{\,km\,s$^{-1}$}
\newcommand{\ip}{SN~2009ip}
\newcommand{\ugc}{UGC~2773 OT2009-1}
\newcommand{\ngc}{NGC~300 OT2008-1}

\def\lsim{\hbox{\rlap{\raise 0.425ex\hbox{$<$}}\lower 0.65ex\hbox{$\sim$}}}
\def\gsim{\hbox{\rlap{\raise 0.425ex\hbox{$>$}}\lower 0.65ex\hbox{$\sim$}}}

\def\arcsec{\hbox{$^{\prime\prime}$}}
\def\arcdeg{\mbox{$^\circ$}}

\newcommand{\hst}{\protect\hbox{\it HST} }

\shorttitle{Diverse Massive Star Outbursts I}
\shortauthors{Foley et~al.}

\begin{document}

 \title{The Diversity of Massive Star Outbursts I: Observations of \ip, \ugc, and Their Progenitors}

\def\cfa{1}
\def\clay{2}
\def\virg{3}
\def\ifa{4}
\def\sao{5}
\def\utah{6}
\def\asu{7}

\author{
{Ryan~J.~Foley}\altaffilmark{\cfa,\clay}, 
{Edo~Berger}\altaffilmark{\cfa},
{Ori~Fox}\altaffilmark{\virg},
{Emily~M.~Levesque}\altaffilmark{\ifa,\sao},
{Peter~J.~Challis}\altaffilmark{\cfa},
{Inese~I.~Ivans}\altaffilmark{\utah},
{James~E.~Rhoads}\altaffilmark{\asu},
{Alicia~M.~Soderberg}\altaffilmark{\cfa}
}
%\email{rfoley@cfa.harvard.edu}

\altaffiltext{\cfa}{
Harvard-Smithsonian Center for Astrophysics,
60 Garden Street, 
Cambridge, MA 02138, USA.
}
\altaffiltext{\clay}{
Clay Fellow. Electronic address rfoley@cfa.harvard.edu .
}
\altaffiltext{\virg}{
Department of Astronomy,
University of Virginia,
P.O. Box 400325,
Charlottesville, VA 22904, USA.
}
\altaffiltext{\ifa}{
Institute for Astronomy,
University of Hawaii,
2680 Woodlawn Dr.,
Honolulu, HI 96822, USA.
}
\altaffiltext{\sao}{
Predoctoral Fellow,
Smithsonian Astrophysical Observatory
}
\altaffiltext{\utah}{
Department of Physics and Astronomy,
University of Utah,
Salt Lake City, UT 84112, USA.
}
\altaffiltext{\asu}{
School of Earth and Space Exploration,
Arizona State University,
P.O. Box 871404,
Tempe, AZ 85287, USA
}

\begin{abstract}
Despite both being outbursts of luminous blue variables (LBVs), \ip\
and \ugc\ have very different progenitors, spectra, circumstellar
environments, and possibly physical mechanisms that generated the
outbursts.  From pre-eruption \hst\ images, we determine that
\ip\ and \ugc\ have initial masses of $\gtrsim 60$ and $\gtrsim 25
M_{\sun}$, respectively.  Optical spectroscopy shows that at peak \ip\
had a 10,000~K photosphere and its spectrum was dominated by narrow H
Balmer emission, similar to classical LBV giant outbursts, also known
as ``supernova impostors.''  The spectra of \ugc, which also have
narrow H$\alpha$ emission, are dominated by a forest of absorption
lines, similar to an F-type supergiant.  Blueshifted absorption lines
corresponding to ejecta at a velocity of 2000 -- 7000~\kms\ are
present in later spectra of \ip\ --- an unprecedented observation for
LBV outbursts, indicating that the event was the result of a
supersonic explosion, rather than a subsonic outburst.  The velocity
of the absorption lines increases between two epochs, suggesting that
there were two explosions in rapid succession.  A rapid fading and
rebrightening event concurrent with the onset of the high-velocity
absorption lines is consistent with the double-explosion model. A
near-infrared excess is present in the spectra and photometry of \ugc\
that is consistent with \about 2100~K dust emission.  We compare the
properties of these two events and place them in the context of other
known massive star outbursts such as $\eta$~Car, \ngc, and SN~2008S.
This qualitative analysis suggests that massive star outbursts have
many physical differences which can manifest as the different
observables seen in these two interesting objects.
\end{abstract}


\defcitealias{Smith09:outburst}{S09}

%%%%%%%%%%%%%%%%%%%%
%%  Introduction  %%
%%%%%%%%%%%%%%%%%%%%

\section{Introduction}\label{s:intro}

Very massive stars appear to go through a phase of instability and
large mass loss; during this stage, a star is a member of the luminous
blue variable (LBV) class (see \citealt{Humphreys94} for a review).
In addition to low-amplitude variability (called S~Dor variability
after the prototypical LBV), where the star ejects mass from its
envelope but its bolometric luminosity remains nearly constant, some
LBVs have ``giant eruptions.''  Giant eruptions can expel $\gtrsim 1
M_{\sun}$ of material, while having a luminosity similar to the
lowest-luminosity supernovae (SNe).  The classical examples of giant
eruptions are P~Cygni in 1600 and $\eta$~Car in 1843, and more recent,
extragalactic examples include SN~1961V (\citealt{Goodrich89,
Filippenko95, VanDyk02}; but see \citealt{Chu04}), V12/SN~1954J
\citep{Smith01, VanDyk05}, SN~1997bs \citep{VanDyk00}, SN~2000ch
\citep{Wagner04}, and V37/SN~2002kg \citep{Weis05, Maund06, VanDyk06}.
Other potential examples exist, but all events listed above have
pre-event imaging where the progenitor star has been identified as a
probable or definite LBV.

Two examples of a new class of stellar eruptions have recently
emerged.  The progenitors of \ngc\ and SN~2008S were both detected in
pre-event {\it Spitzer} images, but were undetected to deep limits in
the optical \citep{Prieto08, Berger09:ngc, Bond09}, indicating
significant reddening from circumstellar dust.  The progenitor stars
were originally believed to have ZAMS masses of 8 -- 20~$M_{\sun}$,
which is below that of the least massive LBVs.  Using the stars in the
vicinity of \ngc, \citet{Gogarten09} found a slightly higher mass
range of 12 -- 25~$M_{\sun}$.  A reasonable range for the initial mass
of the progenitors is \about 10 -- $25 M_{\sun}$, with \ngc, having a
more luminous progenitor, being toward the upper end of that range.

Recently, two transients were discovered with one being very similar
to classical LBV giant eruptions (\ip), and another sharing
characteristics of both LBV giant eruptions and the outbursts of the
dusty stars discussed above (\ugc).  \ip\ was discovered by
\citet{Maza09} in NGC~7259 ($\mu = 32.05 \pm 0.15$~mag\footnote{We use
the distance modulus corresponding to the Hubble distance for NGC~7259
with $H_{0} = 73$~km~s$^{-1}$~Mpc$^{-1}$ and correcting for the Virgo
Infall and Great Attractor flow model of \citet{Mould00}.
\citet{Smith09:outburst} use a distance modulus of 31.55~mag,
which differs by 0.50~mag from our assumed value, corresponding to the
Hubble-flow distance modulus correcting only for the CMB dipole.}; $D
\approx 24$~Mpc) on 2009 August 26 (UT dates will be used throughout
this paper).  \citet{Miller09} noted that \ip\ had been variable for
several years and identified a possible progenitor with $M_{\rm F606W}
\approx -10.1$~mag, in an archival {\it Hubble Space Telescope} (\hst)
image.  The variability and high luminosity of the event led
\citet{Miller09} to suggest that \ip\ was either an LBV or cataclysmic
variable outburst.  An early spectrum of the event showed a blue
continuum with relatively narrow (${\rm FWHM} = 550$~\kms) H Balmer
lines.  The combination of the spectrum with the relatively low
absolute magnitude ($R \approx -13.7$~mag) led us to conclude that
\ip\ was an LBV giant eruption \citep*{Berger09:ip}.  The transient
underwent extreme variability shortly after maximum\footnote{To be
consistent with \citetalias{Smith09:outburst}, we adopt ${\rm MJD} =
55061.5$ and ${\rm MJD} = 55071.75$ as times of maximum light for
\ugc\ and \ip, respectively.  However, we note that the objects may
reach their true maximum later, which \ugc\ already has (as shown by
data presented in \citetalias{Smith09:outburst})}, fading by at least
3~mag in 16~days and rebrightening by 2~mag in the next 10~days
\citep{Li09}, reminiscent of the variability immediately before
maximum of the 1843 eruption of $\eta$~Car \citep{Frew04} and
immediately after maximum in the LBV outburst SN~2000ch
\citep{Wagner04}.  \citet[][hereafter S09]{Smith09:outburst} presented
a historical light curve of \ip\ that begins 5~years before maximum
light, excluding the \hst\ image of the progenitor.  The star varied
by at least 1.5~mag during this time.  \citetalias{Smith09:outburst}
presented additional data which led them to conclude that \ip\ was the
giant eruption of a 50 -- $80 M_{\sun}$ LBV.

\ugc\ was discovered by \citet{Boles09} in UGC~2773 ($\mu = 28.82 \pm
0.17$~mag; $D \approx 6$~Mpc) on 2009 August 18.  It was originally
reported as a possible SN.  We obtained a spectrum and noted that it
had a peculiar spectrum with relatively narrow (${\rm FWHM} =
350$~\kms) H$\alpha$ emission, P-Cygni lines from the \ion{Ca}{2} NIR
triplet, and [\ion{Ca}{2}] emission lines \citep{Berger09:ugc}.  We
also noted that the spectrum was similar to that of \ngc\ and
mentioned that it was possibly a very low-luminosity SN~II or an LBV
outburst, ``but the strong [\ion{Ca}{2}] emission would be unexpected
in this case.''  \citet{Berger09:ugc} also detected a potential
progenitor star in archival \hst\
images. \citetalias{Smith09:outburst} presented a historical light
curve of \ip\ that begins 9~years (excluding the \hst\ image of the
progenitor) before maximum light.  The star slowly increased from
${\rm F814W} = 22.22$~mag to an unfiltered magnitude of 17.70~mag at
maximum light, corresponding to a linear increase of \about
0.4~mag~year$^{-1}$ before outburst.  \citetalias{Smith09:outburst}
concluded from the progenitor identification, their historic light
curve, the peak luminosity, and optical light curves that \ugc\ was
the outburst of a $\gtrsim 20 M_{\sun}$ LBV.

This is the first in a series of papers where we investigate the
diversity of massive star outbursts.  In this paper we demonstrate the
heterogeneity of the class with observations of \ip\ and \ugc.  In
future papers, we will detail the properties of the class and the
links between observations and the physical mechanisms which cause the
outbursts.  In Section~\ref{s:obs}, we present ultraviolet (UV),
optical, and near-infrared (NIR) photometry and optical and NIR
spectroscopy of \ugc\ and \ip.  In this section, we also refine
previous identifications of the progenitors.  In
Section~\ref{s:results}, we examine the progenitor masses, the
spectroscopic characteristics of the outbursts, and the
spectral-energy distributions (SEDs) of the events.  In
Section~\ref{s:disc}, we discuss how these outbursts connect to
previous massive star outbursts and the mass loss history and ultimate
fates of massive stars.  We summarize our conclusions in
Section~\ref{s:con}.

%%%%%%%%%%%%%%%%%%%%
%%  Observations  %%
%%%%%%%%%%%%%%%%%%%%

\section{Observations and Data Reduction}\label{s:obs}

\subsection{Identification and \hst\ Photometry of the Progenitors}

UGC~2773 was observed with \hst\!\!/WFPC2 on 1999 August 14 (Program
8192; PI Seitzer). The observations included two exposures of 600~s
each with the F606W and F814W (roughly $V$ and $I$) filters.  NGC~7259
was observed with \hst\!\!/WFPC2 on 1999 June 29 (Program 6359; PI
Stiavelli).  Exposures of 200 and 400~s were obtained with the F606W
filter.

To determine whether the progenitors of \ip\ and \ugc\ are detected in
the archival \hst\ observations, we performed differential astrometry
using optical observations of the transients.  Observations of \ugc\
were obtained with the Gemini Multi-Object Spectrograph (GMOS) on the
Gemini-North 8-m telescope, and the astrometry was performed using 55
objects in common with the \hst\!\!/WFPC2 images resulting in an
astrometric rms of $\sigma_{\rm GB\rightarrow HST} = 24$~mas in each
coordinate.  Observations of \ip\ were obtained with the Inamori
Magellan Areal Camera and Spectrograph (IMACS) on the Magellan/Baade
6.5-m telescope, and the astrometry was performed using 10 objects in
common with the \hst\!\!/WFPC2 image resulting in an astrometric rms
of $\sigma_{\rm GB\rightarrow HST} = 38$~mas in each coordinate.

The positions of the two transients on the archival \hst\ images are
shown in Figure~\ref{f:proj}.  In both cases, we find a clear
coincidence with objects in the archival \hst\ images.  For \ip\ we
find an offset of $24 \pm 38$~mas relative to the object in the
WFPC2/F606W image, while for \ugc\ we find an offset of $32 \pm
24$~mas relative to the object in the WFPC2/F606W and F814W images.

\begin{figure*}
\begin{center}
\epsscale{1.15}
\plottwo{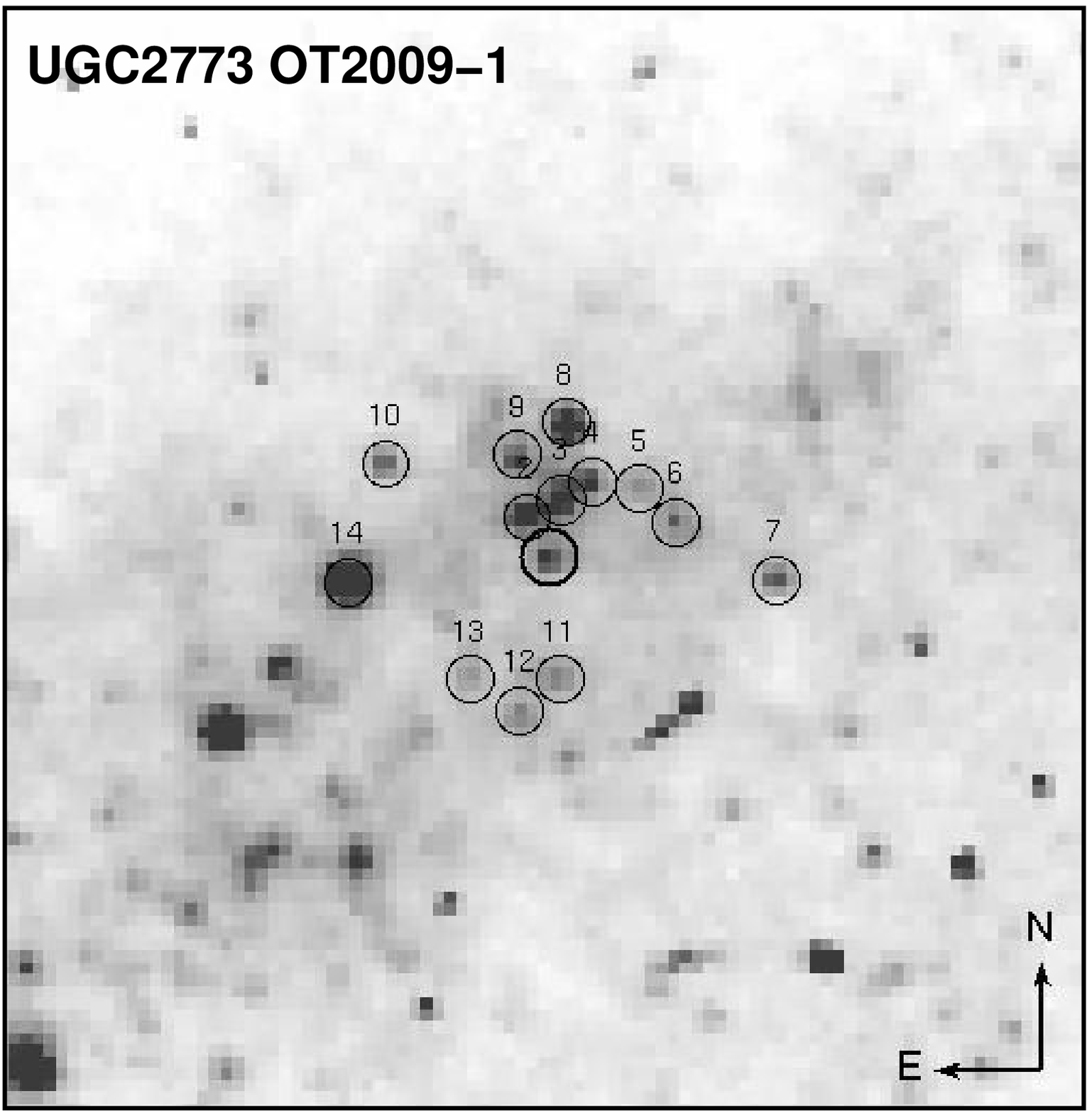}{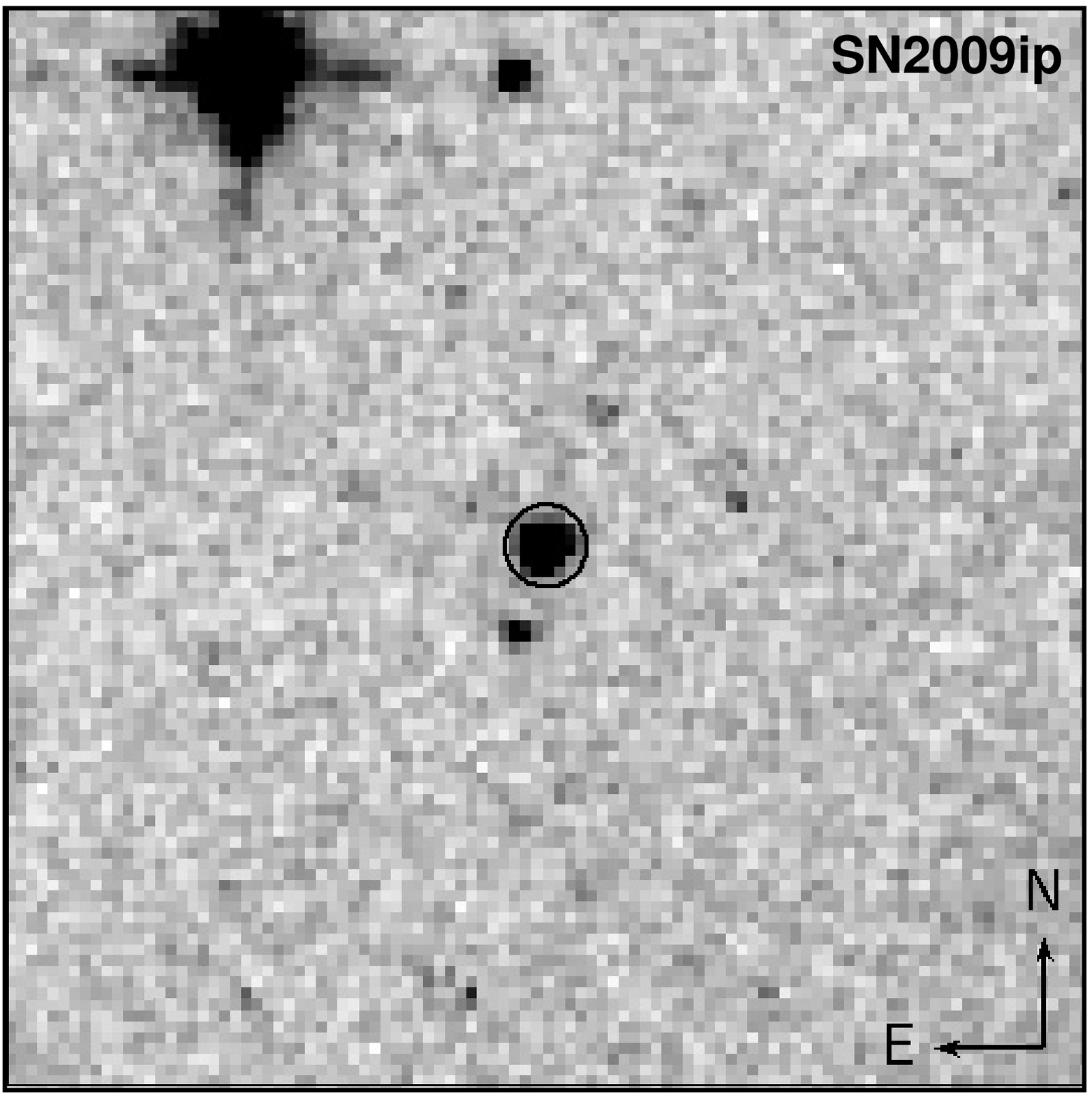}
\caption{{\it HST}/WFPC2 F606W image at the position of \ugc\ (left)
and \ip\ (right) obtained 10~years before maximum.  Both images are
$10\arcsec \times 10\arcsec$, and North is up and East is left.  The
\ugc\ and \ip\ images have pixel scales of 0.1\arcsec~pixel$^{-1}$.
The position of each transient is marked by the black circle whose
radius corresponds to $10 \sigma$ uncertainty in the
position.}\label{f:proj}
\end{center}
\end{figure*}

The measurements of the photometry for \ugc\ and nearby stars were
done using HSTphot 1.1 \citep{Dolphin00}.  HSTphot was run using a
weighted PSF-fit, which is recommended for crowded fields, and a local
sky determination, which is recommended for rapidly varying
backgrounds.  HSTphot performs the conversion from \hst\!\!/WFPC2
flight magnitudes to the Bessel magnitude system.  Our astrometry and
photometry for the nominal progenitor of \ugc\ and nearby stars are
listed in Table~\ref{t:hst}.

\begin{deluxetable*}{lllll}
%\rotate
\tabletypesize{\scriptsize}
\tablewidth{0pt}
\tablecaption{{\it HST} Photometry of Stars Near \ugc\label{t:hst}}
\tablehead{
\colhead{Object} &
\colhead{R.A.} &
\colhead{Dec.} &
\colhead{F606W (mag)} &
\colhead{F814W (mag)}}

\startdata

\phn1\tablenotemark{a} & 03:32:7.240 & +47:47:39.60 & 22.824 (0.032) & 22.286 (0.053) \\
\phn2                 & 03:32:7.258 & +47:47:39.99 & 22.474 (0.027) & 22.164 (0.050) \\
\phn3                 & 03:32:7.229 & +47:47:40.14 & 22.672 (0.036) & 22.222 (0.054) \\
\phn4                 & 03:32:7.203 & +47:47:40.29 & 22.853 (0.035) & 22.591 (0.081) \\
\phn5                 & 03:32:7.162 & +47:47:40.23 & 23.864 (0.072) & 22.729 (0.074) \\
\phn6                 & 03:32:7.132 & +47:47:39.94 & 23.413 (0.051) & 22.748 (0.092) \\
\phn7                 & 03:32:7.048 & +47:47:39.45 & 22.925 (0.034) & 20.956 (0.020) \\
\phn8                 & 03:32:7.225 & +47:47:40.79 & 21.724 (0.016) & 21.043 (0.022) \\
\phn9                 & 03:32:7.266 & +47:47:40.51 & 23.007 (0.039) & 22.548 (0.067) \\
10                    & 03:32:7.377 & +47:47:40.44 & 22.967 (0.035) & 21.682 (0.035) \\
11                    & 03:32:7.230 & +47:47:38.60 & 23.762 (0.092) & 23.182 (0.113) \\
12                    & 03:32:7.263 & +47:47:38.34 & 23.578 (0.058) & 23.095 (0.100) \\
13                    & 03:32:7.306 & +47:47:38.61 & 24.015 (0.081) & 23.304 (0.116) \\
14                    & 03:32:7.409 & +47:47:39.44 & 21.194 (0.026) & 19.934 (0.017)

\enddata

\tablenotetext{a}{Star~1 is identified as the progenitor of \ugc.}

\end{deluxetable*}

We performed photometry of the point source coincident with \ip\ using
a $0.5\arcsec$ aperture and a zero-point of 22.47~mag appropriate for
the F606W filter.  We further applied a correction of $-0.29$~mag to
convert to the Vega system, and applied a correction for the Galactic
extinction of 0.05~mag.  The resulting magnitude of the source is
$21.84\pm 0.25$~mag.  For reasonable colors (see
Section~\ref{sss:ip_mass}), this corresponds to $M_{V} = -10.3$~mag.

\subsection{Ultraviolet, Optical, and Near-Infrared Photometry}

We obtained optical photometry of \ugc\ with the Gemini Multi-Object
Spectrograph (GMOS) on the Gemini-North 8-m telescope in the \gri\
filters.  We performed the photometry using IRAF/{\tt phot} with the
standard GMOS zero-points\footnote{\tt
http://www.gemini.edu/sciops/instruments/gmos/imaging}.  Our results
are presented in Table~\ref{t:phot}.

\begin{deluxetable*}{rlcll}
%\rotate
\tabletypesize{\scriptsize}
\tablewidth{0pt}
\tablecaption{UV and Optical Photometry of \ugc\ and \ip\label{t:phot}}
\tablehead{
\colhead{Object} &
\colhead{MJD} &
\colhead{Filter} &
\colhead{Mag} &
\colhead{Telescope}}

\startdata

\ugc & 51404.13 & F606W   & 22.82 (0.03) & {\it HST} \\
\ugc & 51404.14 & F814W   & 22.29 (0.05) & {\it HST} \\
\ugc & 55078.38 & $J$     & 15.47 (0.06) & Fan Mountain \\
\ugc & 55078.40 & $H$     & 14.99 (0.06) & Fan Mountain \\
\ugc & 55078.39 & $K_{s}$ & 14.63 (0.07) & Fan Mountain \\
\ugc & 55078.53 & $g$     & 18.32 (0.01) & Gemini-North \\
\ugc & 55078.53 & $r$     & 17.22 (0.01) & Gemini-North \\
\ugc & 55078.53 & $i$     & 16.68 (0.01) & Gemini-North \\
\ugc & 55089.31 & $J$     & 15.47 (0.06) & Fan Mountain \\
\ugc & 55089.32 & $H$     & 14.91 (0.07) & Fan Mountain \\
\ugc & 55089.33 & $K_{s}$ & 14.91 (0.09) & Fan Mountain \\
\tableline
\ip  & 51358.50 & F606W   & 21.84 (0.17) & {\it HST} \\
\ip  & 55084.44 & UVW2  & 21.09 (0.19) & {\it Swift} \\
\ip  & 55084.45 & UVM2  & 20.92 (0.28) & {\it Swift} \\
\ip  & 55084.44 & UVW1  & 20.69 (0.18) & {\it Swift} \\
\ip  & 55084.44 & $U$     & 20.29 (0.16) & {\it Swift} \\
\ip  & 55084.44 & $B$     & 20.64 (0.09) & {\it Swift} \\
\ip  & 55084.45 & $V$     & 20.47 (0.37) & {\it Swift}

\enddata

\end{deluxetable*}

We obtained near-infrared (NIR) photometry of \ugc\ with FanCam, a
$1024 \times 1024$ HAWAII-I HgCdTe imaging system on the University of
Virginia's 31-inch telescope at Fan Mountain, just outside of
Charlottesville, VA \citep{Kanneganti09}.  Each epoch consists of
fifteen minutes of integration in \jhks\ bands, which have detection
limits at the $10 \sigma$ level of 0.066, 0.098, and 0.156~mJy (or
18.5, 17.5, and 16.5~mag), respectively.  Individual exposures are
sky-background limited and have an integration time of either 30 or
60~s.  Flat-field frames are composed of dusk and dawn sky
observations.  We employed standard NIR data reduction techniques in
IRAF\footnote{IRAF is distributed by the National Optical Astronomy
Observatory, which is operated by the Association of Universities for
Research in Astronomy (AURA) under cooperative agreement with the
National Science Foundation.}.  Because of the relatively small galaxy
size, it was possible to fit the entire galaxy in a single array
quadrant.  Empty quadrants were efficiently utilized as sky exposures.
Data were taken with the galaxy placed in each quadrant and each
quadrant was reduced separately.  Ultimately, all reduced quadrants
were coadded.  We performed photometry with IRAF's PSF package.  For
magnitude calibration, the transient is compared to 2MASS reference
stars located in the field of view.  Table~\ref{t:phot} lists our
\jhks\ photometry, which is similar to the single epoch \jhks\ data
from \citetalias{Smith09:outburst}.

We obtained UV and optical observations of \ip\ with the {\it Swift}
UV/optical telescope on 2009 September 10.  The data were processed
using standard routines within the HEASOFT package.  Photometry of the
transient in all filters, with the exception of UVW2, was performed
using a 2\arcsec\ aperture to avoid contamination from nearby objects.
Aperture corrections to the standard 5\arcsec\ aperture were
determined using isolated stars; photometry of the source in the UVW2
filter was performed using a 5\arcsec\ aperture.

\subsection{X-ray Observations}

We observed \ip\ and \ugc\ with the {\it Swift} X-ray Telescope on
2009 September 10 for a total exposure time of 9.0 and 4.2~ks,
respectively.  No X-ray counterpart is detected at the position of
either source to a limit of $F_{X} \lesssim 2.8 \times 10^{-14}$ and
$\lesssim 1.1 \times 10^{-13}$~erg~s$^{-1}$~cm$^{-2}$, respectively
($95\%$ limit).  In both cases we assume a power law model with an
electron index of $-2$, and account for the Galactic neutral hydrogen
column.  The corresponding limits on the luminosity are $L_{X}
\lesssim 1.9 \times 10^{39}$ and $\lesssim 4.8 \times
10^{38}$~erg~s$^{-1}$. These limits are comparable to the X-ray
emission from SNe on a similar timescale \citep[e.g.,][]{Soderberg08}.

\subsection{Radio Observations}

We observed the both LBV candidates with the Very Large
Array\footnote{The Very Large Array is operated by the National Radio
Astronomy Observatory, a facility of the National Science Foundation
operated under cooperative agreement by Associated Universities, Inc.}
(VLA) following their optical discovery to search for radio
counterparts, under Rapid Response programs AS1001 and AS1002 (PI
Soderberg).  Our radio observations were carried out at two
frequencies, 8.46 and 22.5 GHz, on dates spanning 2009 September 7.36
- 16.51 in the C-array antenna configuration.  All observations were
taken in standard continuum observing mode with a bandwidth of $2
\times 50$~MHz.  Phase referencing was performed with calibrators
J0325+469 and J2213-254, and we used 3C38 (J0137+331) for flux
calibration.  Data were reduced using standard packages within the
Astronomical Image Processing System (AIPS).

We detect no radio sources in positional coincidence with either
object and derive upper limits summarized in Table~\ref{t:vla}.  At
8.5 GHz, our upper limits\footnote{Upper limits are calculated as the
measured flux density at the optical position summed with $2 \times$
{\it rms} the off-source map noise.} correspond to $L_{\nu} < 1.3
\times 10^{26}$~erg~s$^{-1}$~Hz$^{-1}$ and $L_{\nu} < 2.6 \times
10^{24}$~erg~s$^{-1}$~Hz$^{-1}$ for \ip\ and \ugc, respectively.
These limits are less luminous than an extrapolation of the observed
SN~1961V radio emission at $t \approx 10$~years, to a similarly early
epoch as $L_{\nu} \propto t^{-1.75}$ \citep{Stockdale01}.  We note,
however, that the SN~1961V radio emission may have reached maximum
intensity significantly later than our observations of \ugc\ and \ip,
similar to the radio evolution of SNe~IIn which typically reach
maximum light several years after the explosion
\citep{VanDyk96}.

\begin{deluxetable}{llrr}
\tablecaption{VLA observations of \ugc\ and \ip\label{t:vla}}
\tablewidth{0pt}
\tablehead{
\colhead{Object} & \colhead{Date} & \colhead{$F_{\nu, 8.5}$} & \colhead{$F_{\nu, 22}$} \\
\colhead {} & \colhead{(UT)} & \colhead{($\mu$Jy)} & \colhead{($\mu$Jy)}
}

\startdata

\ip     & 2009 Sep 7.36  & \nodata     & $-122 \pm 243$ \\
\nodata & 2009 Sep 9.24  & $67 \pm 47$ & \nodata \\
\ugc    & 2009 Sep 13.57 & \nodata     & $-57 \pm 66$ \\
\nodata & 2009 Sep 16.51 & $11 \pm 27$ & \nodata

\enddata

\tablecomments{Uncertainties are $1\sigma$ {\it rms} map noise.}

\end{deluxetable}

A comparison of these radio upper limits for the outbursts to the
observed properties of other core-collapse SNe places them among the
least luminous events, 2-4 orders of magnitude less luminous than the
most powerful SNe~IIn, and 4-200 times higher than the early radio
signal seen for SN~1987A \citep{Ball95}.  Through this simple
comparison we emphasize that radio data alone cannot distinguish
between massive star outbursts and catastrophic explosions.  However,
with the 10-fold increase in continuum sensitivity provided by the
EVLA we will begin to map out the radio properties for massive star
outbursts and enable direct comparisons with those of core-collapse
SNe (e.g., NRAO Key Project AS1020, "Exotic Explosions,Eruptions,and
Disruptions: A New Transient Phase-Space", PI Soderberg).

\subsection{Optical Spectroscopy}\label{ss:opt_spec}

We obtained low- and medium-resolution spectra of \ip\ and \ugc\ with
the MagE spectrograph \citep{Marshall08} on the Magellan Clay 6.5~m
telescope, the Blue Channel spectrograph \citep*{Schmidt89} on the MMT
6.5~m telescope, and GMOS \citep{Hook04} on the Gemini-North 8~m
telescope.  A journal of our optical spectroscopic observations can be
found in Table~\ref{t:spec}.

\begin{deluxetable*}{lllcrl}
%\rotate
\tabletypesize{\scriptsize}
\tablewidth{0pt}
\tablecaption{Log of Optical Spectral Observations\label{t:spec}}
\tablehead{
\colhead{} &
\colhead{} &
\colhead{Telescope /} &
\colhead{Grating /} &
\colhead{Exposure} &
\colhead{} \\
\colhead{Phase\tablenotemark{a}} &
\colhead{UT Date} &
\colhead{Instrument} &
\colhead{Central Wavelength (\AA)} &
\colhead{(s)} &
\colhead{Observer\tablenotemark{b}}}

\startdata

\cutinhead{\ugc} 

15.1 & 2009 Sep.\ 2.6  & Gemini/GMOS      & R400/7000 & $2 \times 1200$      & KO, RM \\
32.9 & 2009 Sep.\ 20.4 & MMT/Blue Channel & 300/5787  & $3 \times 1200$      & PC \\
33.9 & 2009 Sep.\ 21.4 & MMT/Blue Channel & 300/5787  & 1200                 & PC \\
33.9 & 2009 Sep.\ 21.4 & MMT/Blue Channel & 832/4029  & $2 \times 900$       & PC \\
34.0 & 2009 Sep.\ 21.5 & MMT/Blue Channel & 832/4830  & $2 \times 900$       & PC \\
34.0 & 2009 Sep.\ 21.5 & MMT/Blue Channel & 832/6563  & $3 \times 900$       & PC \\
95.8 & 2009 Nov.\ 22.3 & MMT/Blue Channel & 300/5787  & $2 \times 1800$      & PC \\

\cutinhead{\ip}

 3.5 & 2009 Sep.\ 1.3  & Clay/MagE        & \nodata   & $3 \times 900$       & II \\
22.5 & 2009 Sep.\ 20.3 & MMT/Blue Channel & 300/5787  & $4 \times 1200$      & PC \\
23.5 & 2009 Sep.\ 21.2 & MMT/Blue Channel & 832/4029  & $2 \times 1200$      & PC \\
23.5 & 2009 Sep.\ 21.3 & MMT/Blue Channel & 832/4830  & $2 \times 1200$      & PC \\
23.5 & 2009 Sep.\ 21.3 & MMT/Blue Channel & 832/6563  & 1200                 & PC \\
85.6 & 2009 Nov.\ 22.1 & MMT/Blue Channel & 300/5787  & $3 \times 1200$      & PC

\enddata

\tablenotetext{a}{Days since maximum, MJD 55,061.5 and 55,071.8 for
\ugc\ and \ip\ \citepalias{Smith09:outburst}, respectively.}

\tablenotetext{b}{II = I.\ Ivans, KO = K.\ Olsen, PC = P.\ Challis, RM
= R.\ McDermid}

\end{deluxetable*}

Standard CCD processing and spectrum extraction were accomplished with
IRAF.  The data were extracted using the optimal algorithm of
\citet{Horne86}.  Low-order polynomial fits to calibration-lamp
spectra were used to establish the wavelength scale, and small
adjustments derived from night-sky lines in the object frames were
applied.  For the MagE spectra, the sky was subtracted from the images
using the method described by \citet{Kelson03}.  The GMOS data were
reduced using the Gemini IRAF package \citep[for details, see
][]{Foley06}.  We employed our own IDL routines to flux calibrate the
data and remove telluric lines using the well-exposed continua of the
spectrophotometric standards \citep{Wade88, Foley03, Foley09:08ha}.

Representative spectra of \ip\ and \ugc\ are presented in
Figure~\ref{f:opt_spec}.  Both objects have similar blue continua, but
the line features are very different.  \ip\ has few line features
besides strong H Balmer lines, Na~D, and \ion{He}{1}.  Although \ugc\
has a strong H$\alpha$ line, it is much narrower and weaker than that
of \ip.  \ugc\ also displays many additional narrow line features,
including lines from intermediate mass and Fe-group elements.
Blueward of \about 5500~\AA, \ugc\ is dominated by a forest of
\ion{Fe}{2} lines.  Finally, \ugc\ has very strong \ion{Ca}{2} NIR
triplet lines and [\ion{Ca}{2}] $\lambda\lambda 7291$, 7324 lines.
These features, and particularly [\ion{Ca}{2}] are rarely seen in
classical LBV outbursts, but were distinguishing features of SN~1999bw
\citep{Garnavich99}, SN~2008S \citep[e.g.,][]{Smith09:08s}, and \ngc\
\citep[e.g.,][]{Berger09:ngc}.

\begin{figure*}
\begin{center}
\epsscale{0.6}
\rotatebox{90}{
\plotone{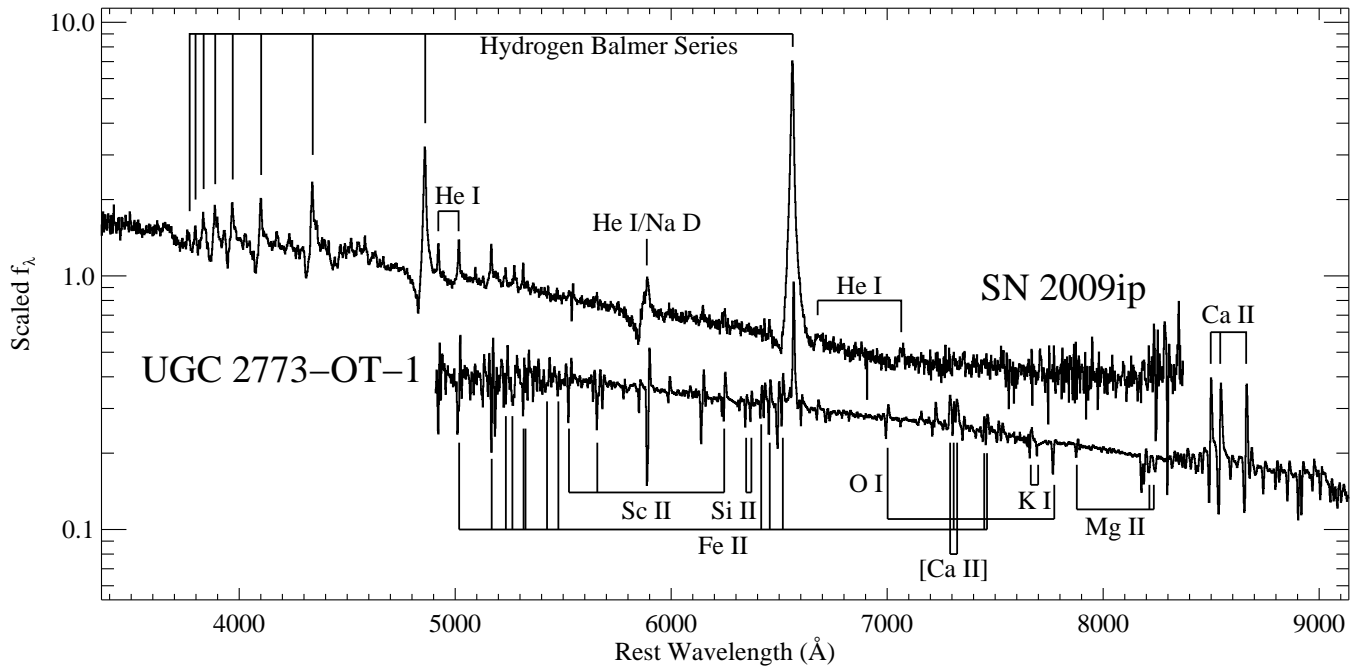}}
\caption{Optical spectra of \ip\ and \ugc.  The spectrum of \ugc\ has
been dereddened by $E(B-V) = 0.564$~mag.  Several lines are identified
and marked.}\label{f:opt_spec}.
\end{center}
\end{figure*}

The spectra from 2009 September 21 (corresponding to days 34 and 24
for \ugc\ and \ip, respectively) were obtained on the same night as
the LRIS spectra shown by \citetalias{Smith09:outburst}.

\subsection{Near-Infrared Spectroscopy}

On 2009 September 9 (22~days after maximum), we obtained a 2400~s NIR
spectrum of \ugc\ with TripleSpec, a medium resolution NIR
spectrograph located at Apache Point Observatory.  This spectrograph
is one of three NIR, cross-dispersed spectrographs covering
wavelengths from 1 -- 2.4~\micron\ simultaneously at a resolution of
\about 3500 \citep{Wilson04, Herter08}.  We collected eight, 300-s
sky-background limited exposures, for a total integration time of
2400~s.  We extracted the spectrum with a modified version of the
IDL-based SpexTool \citep*{Cushing04}.  This tool removes any
contribution from the underlying galactic arm by fitting the
background with a 2nd order polynomial.

%%%%%%%%%%%%%%%
%%  Results  %%
%%%%%%%%%%%%%%%

\section{Results}\label{s:results}

\subsection{Progenitor Masses}

\subsubsection{\ip}\label{sss:ip_mass}

We use the absolute magnitude, $M_{V} = -10.3$~mag, of the progenitor
of \ip, along with an estimated range of $V - I$ colors of $-0.05$ to
$1.4$~mag (representative of LBV colors spanning from O to F spectral
types), to plot it as a line on a color-magnitude diagram
(Figure~\ref{f:cmd}).  For this progenitor we find a Milky Way
extinction of $E(B-V) = 0.019$~mag ($A_{V} = 0.05$ mag;
\citealt*{Schlegel98}).  We adopt a distance modulus of $\mu =
32.05$~mag for NGC~7259 (see discussion in Section~\ref{s:intro}), and
assume no additional host galaxy or circumstellar extinction.

\begin{figure}
\begin{center}
\epsscale{1.23}
\plotone{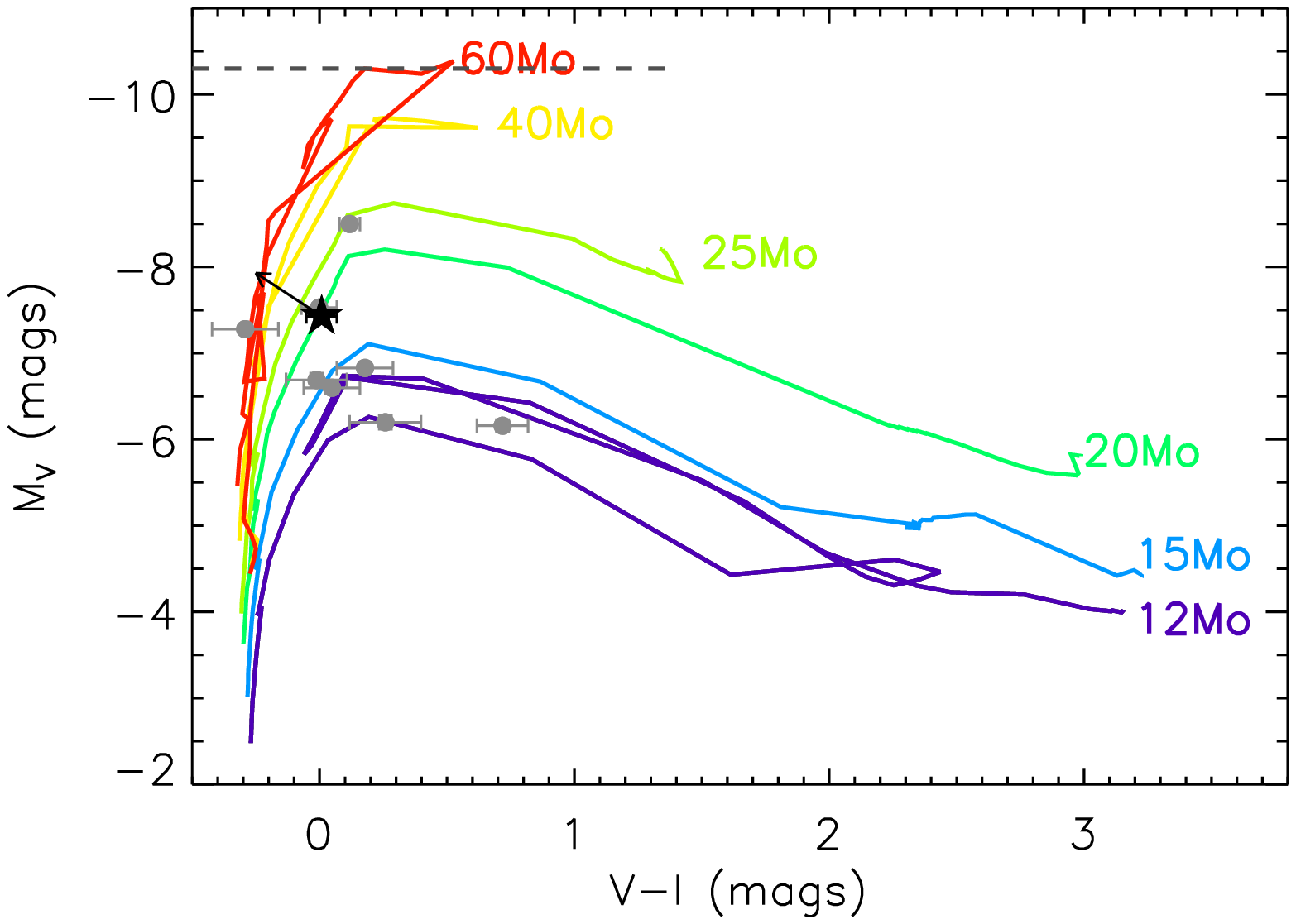}
\caption{Color-magnitude diagram ($V-I$ vs.\ $M_{V}$) for the
progenitor of \ugc\ (star) and stars spatially located within the same
star cluster (grey circles).  The measurements have been corrected for
the Milky Way extinction of $A_{V} = 1.75$~mag and $E(V-I) =
0.902$~mag, but no host or circumstellar extinction is assumed for the
stars.  For comparison, solar metallicity, non-rotating, ``standard''
mass loss stellar evolution tracks are also plotted
\citep{Schaller92}.  The progenitor of \ugc\ has the same colors and
absolute magnitude of a $20 M_{\sun}$ model.  The stars in the cluster
are consistent with the models of stars with ZAMS masses $\leq 25
M_{\sun}$, but a single star is also consistent with a much higher
mass.  The arrow represents $A_{V} = 0.5$~mag of additional extinction
(assuming $R_{V} = 3.1$).  The dashed line represents the absolute
magnitude of the progenitor of \ip\ with a reasonable range of
colors.  The luminosity of the progenitor of \ip\ is consistent with
an initial mass of $\gtrsim 60 M_{\sun}$.}\label{f:cmd}
\end{center}
\end{figure}

In Figure~\ref{f:cmd}, we compare the color of the \ip\ progenitor to
the non-rotating, standard mass-loss evolutionary tracks of the Geneva
group \citep{Schaller92}.  From this plot we can place a lower initial
mass limit of $60 M_{\sun}$ on the progenitor of \ip\ in the absence
of a color estimate for this progenitor, the higher-mass evolutionary
tracks all coincide with its estimated location on the color-magnitude
diagram, precluding us from placing an upper limit on this initial
mass estimate.  Figure~\ref{f:cmd} assumes a solar metallicity for
these tracks; however, we find that our progenitor mass prediction is
consistent across the full range of metallicities accommodated by the
Geneva evolutionary tracks ($Z = 0.05 Z_{\sun}$ to $Z = 2 Z_{\sun}$).
It should be noted that an increased amount of extinction, from the
host galaxy or circumstellar environment, could also effectively
increase the estimated initial mass of this progenitor.

\citetalias{Smith09:outburst} estimated the initial mass for the
progenitor of \ip\ to be 50 -- $80 M_{\sun}$.  Although the {\it HST}
photometry from \citetalias{Smith09:outburst} is the same as that
presented here, their assumed distance modulus is 0.50~mag smaller
than our assumed value.  They also make no color correction to
transform the F606W measurements into $V$.  Despite these differences,
the two mass ranges are similar.

\subsubsection{\ugc}

Using its $M_{V}$ and $V-I$ color, we are able to determine an
estimate of the initial mass for the progenitor of \ugc\
(Figure~\ref{f:cmd}).  There is significant Milky Way extinction of
$E(B-V) = 0.564$~mag ($A_{V} = 1.75$~mag; \citealt*{Schlegel98}),
which we convert to $E(V-I) = 0.902$~mag \citep{Schultz75}.  We use a
distance modulus of $\mu = 28.82$~mag and initially assume no host
galaxy or circumstellar extinction.

From Figure~\ref{f:cmd}, we find that the progenitor of \ugc\ is
consistent with an initial mass of \about $20 M_{\sun}$. Our
progenitor mass prediction remains the same across the full range of
metallicities covered by the Geneva evolutionary tracks, and is
consistent with the value found by \citetalias{Smith09:outburst}.

We have also performed this procedure on several stars in the vicinity
of the progenitor of \ugc.  Assuming that all of these stars are part
of a cluster and were formed at the same time, they should place
additional limits on the current maximum-mass stars of the cluster.
These stars are all consistent with an initial mass of $M_{\sun}
\lesssim 25 M_{\sun}$.  There is a single star that is particularly
blue (and therefore potentially very massive), but it is still
consistent with an initial mass of $25 M_{\sun}$.  The likely
association of the progenitor of \ugc\ with this cluster and its upper
mass limit of \about 25~$M_{\sun}$ further supports the initial mass
estimate for the progenitor of \ugc.

Considering the blue colors of the stars in the cluster, it is
unlikely that they are significantly reddened by host galaxy dust.  As
shown in Figure~\ref{f:cmd}, a relatively small amount of extinction
could significantly increase our initial mass estimate for \ugc.  In
Section~\ref{ss:sed} we show that there was likely a significant
amount of circumstellar dust existing before the outburst, indicating
that the progenitor had an initial mass much larger than the
reddening-free estimate of $20 M_{\sun}$.

The combination of the reddening-free initial mass estimate for the
progenitor of \ugc, the initial mass estimates of stars likely within
the same cluster as the progenitor, and the probably circumstellar
dust extinction give us a conservative lower limit on the initial mass
of the progenitor of \ugc\ of \about $25 M_{\sun}$.

\subsection{Spectroscopic Comparisons}

\subsubsection{\ip}\label{sss:ip_spec}

We present the 24 and 86~day spectra of \ip\ in
Figure~\ref{f:ip_spec}.  In the upper panel of Figure~\ref{f:ip_spec},
the 24~day spectrum is compared to the 2~day spectrum of the LBV
outburst SN~1997bs \citep{VanDyk00}.  Both objects have blue continua,
strong and narrow H Balmer lines, and \ion{He}{1} and \ion{Fe}{2}
emission lines.  Unlike SN~1997bs, \ip\ has a particularly strong
\ion{He}{1} $\lambda 5876$ line (with some possible contribution from
Na~D), and all H Balmer lines and \ion{He}{1} $\lambda 5876$ show
strong absorption features with minima blueshifted by \about
3000~\kms\ (see Section~\ref{sss:halpha} for a detailed discussion of
this high-velocity absorption).

\begin{figure}
\begin{center}
\epsscale{0.9}
\rotatebox{90}{
\plotone{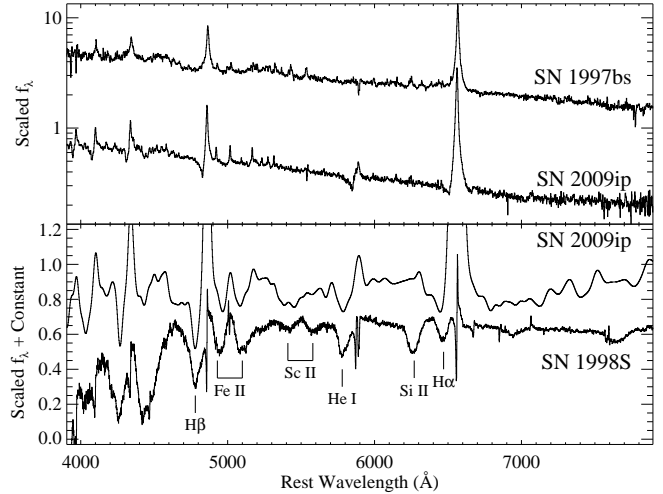}}
\caption{Top panel: the 23~day optical spectrum of \ip\ compared to
the 2~day spectrum of SN~1997bs \citep{VanDyk00}.  Bottom panel: the
86~day optical spectrum of \ip\ after smoothing and subtracting a
10,000~K blackbody (see text for details).  For comparison, the 25~day
spectrum of SN~1998S (after subtracting a 10,000~K blackbody) is also
shown \citep{Leonard00}.  Prominent, high-velocity lines have been
marked.}\label{f:ip_spec}
\end{center}
\end{figure}

Inspecting the 25~day spectrum from \citetalias{Smith09:outburst}, we
see some indication of the 3000~\kms\ absorption component,
particularly for H$\beta$ and \ion{He}{1} $\lambda 5876$; however, the
absorption is much stronger in the spectra presented here (which were
obtained at very similar times).  We have reduced our spectra with
many different extraction regions and backgrounds, with the
high-velocity absorption features are present in all reductions.  The
same feature is present in all spectra taken with the MMT on days 23
and 24, which were taken with different gratings and wavelength
regions.  It is also present on day 86, but with a different velocity.
The absorption is present for all Balmer lines and \ion{He}{1}
$\lambda 5876$.  Despite the apparent differences with the concurrent
Keck spectrum \citepalias{Smith09:outburst}, we are confident that the
high-velocity absorption features are real and not an artifact of the
spectral reductions.

The bottom panel of Figure~\ref{f:ip_spec} displays the spectra of
\ip\ and SN~IIn~1998S (after subtracting a 10,000~K blackbody spectrum
from both) from 86 and 25~days, respectively.  An inverse-variance
weighted Gaussian filter (with a width of 1000~\kms) has been applied
to the spectrum of \ip\ \citep{Blondin06}.  This filtering will smear
out features with intrinsic widths less than 1000~\kms, but will
appropriately smooth features on larger scales.  The high-velocity
absorption in the 86~day spectrum of \ip\ is at a {\it higher}
velocity than at 24~days.  At this epoch, the velocity of the
fast-moving \ip\ ejecta are very similar to that of SN~1998S.
Although the H Balmer emission lines are much stronger in \ip, most
other features are similar in the two spectra.  In particular, \ip\
shows the H Balmer, \ion{He}{1}, \ion{Sc}{2}, and \ion{Fe}{2} features
seen in SN~1998S.  \ip\ is missing the strong absorption at 6250~\AA\
that is attributed to \ion{Si}{2} in SN~1998S \citep{Leonard00}.  This
feature may be the result of a significant amount of nuclear burning,
and thus not present in the ejecta of \ip.

\subsubsection{\ugc}

As discussed in Section~\ref{ss:opt_spec}, \ugc\ has a spectrum with
narrow H$\alpha$ emission, [\ion{Ca}{2}] emission, and P-Cygni
absorption from many intermediate-mass and Fe-group elements.  Perhaps
the most distinguishing feature compared to other massive star
outbursts is the [\ion{Ca}{2}] emission.  In Figure~\ref{f:ugc_spec},
we compare the 15~day spectrum of \ugc\ to spectra of the
low-luminosity transients \ngc\ \citep{Berger09:ngc} and SN~2008S
\citep{Smith09:08s}, as well as SN~IIn~1994W \citep{Chugai04}; all of
these objects have [\ion{Ca}{2}] emission in their spectra.

\begin{figure}
\begin{center}
\epsscale{0.9}
\rotatebox{90}{
\plotone{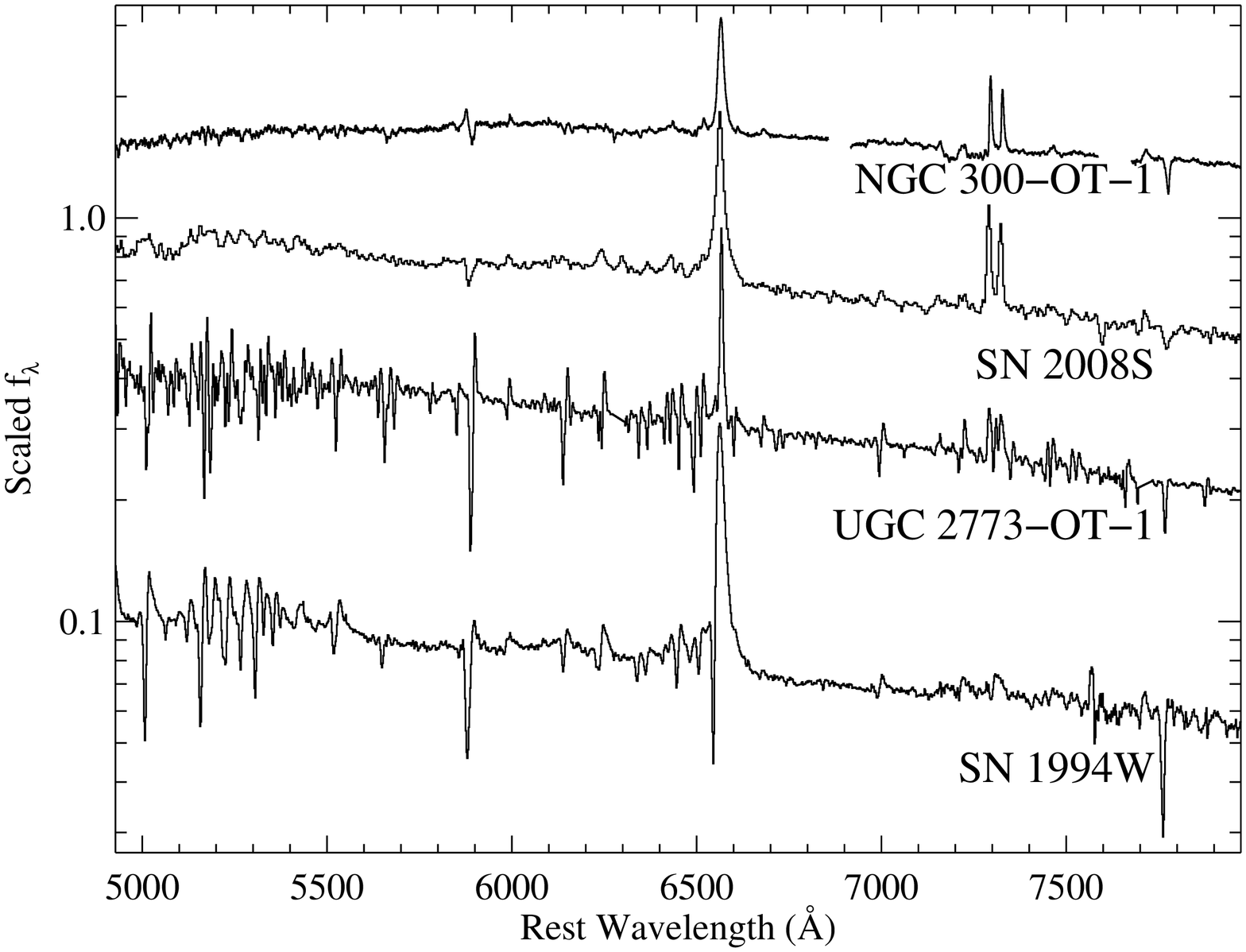}}
\caption{Optical spectra of \ugc, \ngc\ \citep{Berger09:ngc}, SN~1994W
\citep{Chugai04}, and SN~2008S \citep{Smith09:08s}.  All spectra have
narrow H$\alpha$ and [\ion{Ca}{2}] emission; however, \ngc\ and
SN~2008S lack the forest of lines (especially \ion{Fe}{2}) that \ugc\
and SN~1994W display.}\label{f:ugc_spec}
\end{center}
\end{figure}

All spectra in Figure~\ref{f:ugc_spec} are relatively similar.  The
continuum of each spectrum is well-described by a blackbody spectrum,
with all four objects having a similar temperature.  Each object has a
prominent H$\alpha$ emission line, with \ugc\ having a narrower line
than the other objects.  Additionally, SN~1994W has a strong H$\alpha$
absorption line blueward of its emission peak.

\ngc\ and SN~2008S are very similar objects with massive (10 --
25~$M_{\sun}$), dusty progenitors \citep{Prieto08, Berger09:ngc,
Bond09}.  Their spectra share many characteristics with the yellow
hypergiant IRC+10240 \citep{Smith09:08s}.  Although \ugc\ shares some
spectroscopic properties with these two transients and IRC+10240 (see
\citetalias{Smith09:outburst} for additional discussion), the latter
objects lack the forest of absorption lines in \ugc.  These lines are
reminiscent of an F-type supergiant.  The P-Cygni profiles of these
lines and the hydrogen Balmer emission are very similar to S~Dor
during a cool phase \citep[e.g.,][]{Massey00}.

SN~1994W was very luminous at peak ($M_{V} \approx -19$~mag), but
generated at most 0.03~$M_{\sun}$ of $^{56}$Ni \citep*{Sollerman98}.
\citet{Dessart09} presented an alternative method of producing the
photometric and spectroscopic properties of this object: the
collision of two massive hydrogen shells ejected from the star with
no core collapse.  Spectra of SNe~IIn are rather heterogeneous (see
Figure~5 of \citealt{Smith09:06gy} for a comparison of various
objects), and SN~1994W is relatively distinct for its narrow
absorption features.  Given the spectral similarity between \ugc\ and
SN~1994W, the strict upper limit of $^{56}$Ni mass in SN~1994W, and
the alternative model of \citet{Dessart09}, one must further question
if SN~1994W destroyed its progenitor star.

\subsubsection{Contrasting \ip\ and \ugc}

At $t = 0$~days, the temperature of \ugc\ is \about 7000~K (see
Section~\ref{ss:sed}), which is similar to the temperature during the
``eruptive'' state of LBVs \citep[e.g.,][]{Humphreys94}.  This
contrasts with the higher temperature of 10,000~K derived for \ip\
(see Section~\ref{ss:sed}), which lacks the narrow Fe-group absorption
features.  Many other LBV giant eruptions have temperatures similar to
that of \ip\ \citep{Humphreys94}.  \ip\ was \about 2~mag brighter at
peak than \ugc, and \ip\ had a much larger increase in luminosity
during the year before maximum than \ugc, increasing by $\gtrsim
5$~mag and \about 1~mag over one year, respectively
\citepalias{Smith09:outburst}.  The fast-moving ejecta of \ip\ also
contrasts with the relatively slow outflow of \ugc.

The photometric and spectroscopic differences of these objects
suggests different physical mechanisms.  Clearly a supersonic
explosion is necessary to produce the high-velocity absorption
features of \ip, while \ugc\ shows no indication of an explosion.  The
differences in temperature and luminosity increase are also indicative
of more energy injection (per unit mass) for \ip.  A plausible
explanation is that \ip\ is a LBV giant eruption triggered by an
explosion, while \ugc\ is a particularly luminous S~Dor eruption.

\subsection{Line Profiles}

In this section, we examine the line profiles of H$\alpha$ and Ca
lines.  These features provide an indication of the kinematics of the
emitting material.  The narrow lines are a tracer of the pre-shock
circumstellar material, while the high-velocity absorption features in
the spectra of \ip\ probe the outburst ejecta.

\subsubsection{H$\alpha$}\label{sss:halpha}

In Figure~\ref{f:halpha}, we present the H$\alpha$ line profiles of
\ugc\ and \ip.  Three separate epochs are shown for each object.  Both
objects have asymmetric line profiles.  There are absorption
components at about $-350$\kms\ for \ugc\ and between $-3000$ and
$-6000$~\kms\ for \ip.  The line profile of \ip\ has a different shape
and is much broader than that of \ugc.  We have attempted to fit these
line profiles, but because of the asymmetry of the profiles, we first
fit only the redshifted portion of each line profile and then add an
absorption component to reproduce the blueshifted profile.

\begin{figure}
\begin{center}
\epsscale{1.}
\rotatebox{90}{
\plotone{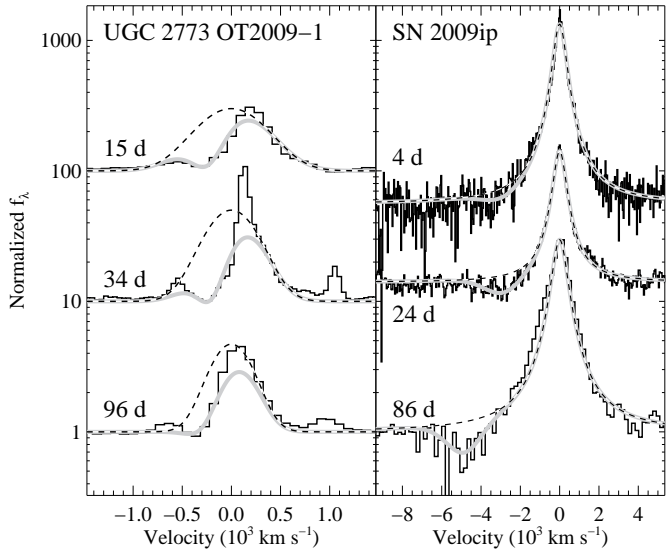}}
\caption{Normalized spectra of \ugc\ and \ip\ near H$\alpha$.  The
line profiles are fit with Gaussian and Lorentzian profiles,
respectively.  A profile fit to the redshifted portion of each profile
is shown as a dashed line.  The grey lines correspond to the
redshifted profile with a Gaussian absorption component added.  Narrow
[\ion{N}{2}] can be seen in the spectrum of \ugc.}\label{f:halpha}
\end{center}
\end{figure}

For the 15~day spectrum of \ugc, we fit a Gaussian with ${\rm FWHM} =
780$~\kms\ to the red side of the feature.  This value is twice that
of the value found by \citetalias{Smith09:outburst} for a spectrum
from day 22, but examination of their figures suggest that they
reported half-width at half maximum (HWHM) or the standard deviation
of the Gaussian fit (which is smaller by a factor of 2.35) rather than
FWHM.  The 34~day spectrum of \ugc\ is contaminated by host-galaxy
emission lines, making a fit to the inner regions of the line profile
problematic.  Ignoring this region, we were able to fit the redshifted
portion of the line profile with a single Gaussian with ${\rm FWHM} =
590$~\kms.  The 96~day spectrum has lower resolution, but is
successfully fit by a Gaussian profile with ${\rm FWHM} = 470$~\kms.

To account for the asymmetric profile, we add an absorption component
to the Gaussian line profiles.  Fitting the full profile with two
Gaussian functions, the emission component fit to the red side of the
line and the absorption component added to fit the blue side of the
line, we find absorption minima at $-180$, $-110$, and $-80$~\kms\ for
the 15, 34, and 96~day spectra, respectively.  This is different from
the value of the actual minimum ($-350$~\kms) since the relatively
strong emission masks the true minimum.

The line profiles of the first two spectra (days 4 and 24) of \ip\ are
well fit by Lorentzian profiles with ${\rm FWHM} = 780$~\kms\ and the
third is best fit by a Lorentzian profile with ${\rm FWHM} =
890$~\kms, which are larger than that found by
\citetalias{Smith09:outburst}, 550~\kms.  A Lorentzian profile of
550~\kms\ is not a particularly bad fit to our data, but we find that
the larger velocities better represent the data.  One can also see in
Figure~8 of \citetalias{Smith09:outburst}, that the 550~\kms\
Lorentzian slightly underpredicts the true FWHM of the line, so the
data appear to be consistent.

In the 24~day spectrum of \ip, we see an absorption feature with a
minimum at a velocity of about $-3000$~\kms. (This high-velocity
absorption is seen for all Balmer lines with varying instrument
configurations and on two epochs; see Section~\ref{sss:ip_spec}.)
This feature is well fit by including a Gaussian absorption component
with a minimum at $-2800$~\kms.  Adding a component with this velocity
also improves the fit to the 4~day H$\alpha$ profile slightly, but not
in a significant way.  The 86~day spectrum shows an even {\it
stronger} high-velocity absorption component with the minimum of the
absorption at a {\it larger} velocity of $-4800$~\kms.  The blue wing
of the absorption component, representing the fastest moving material,
corresponds to a velocity of about $-4500$ and $-7000$~\kms\ for the
24 and 86~day spectra, respectively.

These velocities are significantly larger than the outflow velocity of
550~\kms\ assumed by \citetalias{Smith09:outburst}.  They are much
larger than the wind speed of LBVs and are larger than the measured
velocity for any LBV eruption with the exception of the 1843 eruption
of $\eta$~Car, which had some material expelled at 3000 -- 6000~\kms\
\citep{Smith08:car}.  The velocities measured for \ip\ are similar to
that of the ejecta of typical core-collapse SNe (such as SN~1998S; see
Figure~\ref{f:ip_spec} and Section~\ref{sss:ip_spec}) and are somewhat
similar to that of Wolf-Rayet winds \citep[e.g.,][]{Abbott87}.  We
discuss the implications of these features in Section~\ref{ss:exp}.

\subsubsection{Permitted and Forbidden \ion{Ca}{2}}

Only our first spectrum of \ip\ covers the \ion{Ca}{2} NIR triplet,
and no spectrum shows obvious [\ion{Ca}{2}] $\lambda\lambda 7291$,
7325 lines, similar to the spectra presented by
\citetalias{Smith09:outburst}.  Furthermore, the Ca H\&K lines are
confused by the strong Balmer sequence in \ip.  Because of these
factors, it is difficult to evaluate the characteristics of the
\ion{Ca}{2} behavior in this object (other than the absent
[\ion{Ca}{2}] lines).

\ugc, on the other hand, has strong \ion{Ca}{2} features.  This can be
seen in Figure~\ref{f:opt_spec}.  We examine the Ca H\&K,
[\ion{Ca}{2}] $\lambda\lambda 7291$, 7325, and \ion{Ca}{2} NIR triplet
line profiles in Figure~\ref{f:ca2}.  The Ca H\&K lines show a broad
absorption extending from $-1000$ to $+500$~\kms\ and a minimum at
about $-50$~\kms\ that does not appear to change significantly between
the two epochs.  Each component of the \ion{Ca}{2} NIR triplet shows a
strong P-Cygni profile with a minimum at approximately $-250$~\kms,
slightly larger than the minima of Ca H\&K.

\begin{figure}
\begin{center}
\epsscale{3.}
\rotatebox{90}{
\plotone{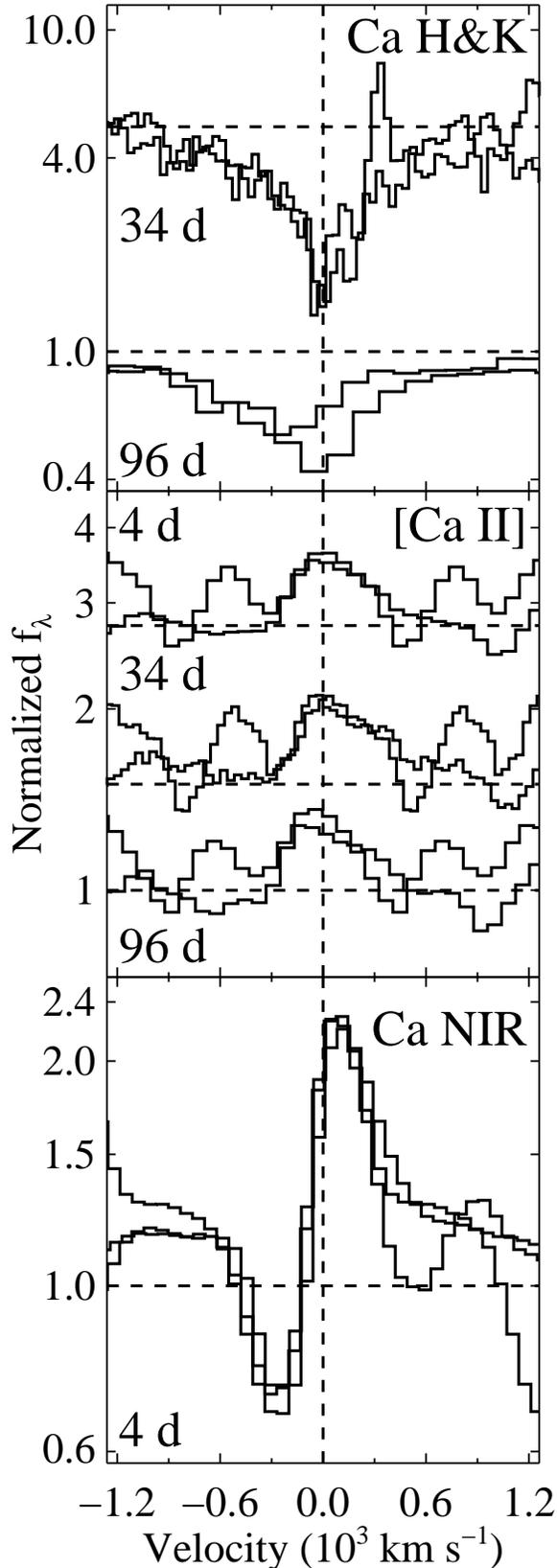}}
\caption{Normalized spectra of \ugc\ near Ca H\&K (top),
[\ion{Ca}{2}] $\lambda\lambda 7291$, 7325 (middle), and \ion{Ca}{2}
NIR triplet (bottom).  Dashed lines indicate the continuum flux and
zero velocity for each line.  Each member of the multiplet is
overplotted for a given spectrum.}\label{f:ca2}
\end{center}
\end{figure}

The [\ion{Ca}{2}] $\lambda\lambda 7291$, 7325 lines are visible in all
epochs of our spectroscopy.  We confirm the additional line between
this doublet seen by \citetalias{Smith09:outburst} and identify this
as \ion{Fe}{2} $\lambda 7308$.  Both [\ion{Ca}{2}] lines have
asymmetric profiles in all spectra; the peak is at zero velocity, but
the emission extends further to the red than to the blue.  The lines
from all epochs have FWHMs of \about 400~\kms, which is about half the
width of H$\alpha$ (see Section~\ref{sss:halpha}), similar to that
found for \ngc\ \citep{Berger09:ngc}.

\subsection{Spectral Energy Distribution and Dust Emission}\label{ss:sed}

Using our available photometry and spectroscopy, we can examine the
spectral energy distribution (SED) of both objects.  We have only
optical spectra of \ip, which limits our ability to examine multiple
blackbody components for this object.  A 10,000~K blackbody fits our
optical spectra well, which is consistent with that found by
\citetalias{Smith09:outburst}.

Our single epoch of {\it Swift} photometry occurred during the
dramatic fading of the light curve immediately following maximum
brightness \citepalias{Smith09:outburst}.  In Figure~\ref{f:ipbb}, the
{\it Swift} photometry is combined with the unfiltered photometry
(approximately $R$ band) of \citetalias{Smith09:outburst} (with an
uncertainty of 0.5~mag to account for the 16~hour difference in the
epoch of the observations) during the minimum.  We overplot the 23~day
spectrum for comparison.  The optical photometry is consistent with
the optical spectrum and a 10,000~K blackbody.  The UVW2 flux is also
consistent with this blackbody, however, the UVM2 and UVW1
measurements fall well below this curve.  Although this may be the
result of line blanketing, these data are also consistent with a
blackbody curve with a temperature as low as 8000~K.  If we ignore the
UVW2 measurement, the data can be fit by a 7000~K blackbody.  Although
our data suggest a possible change in the SED during the fading event,
the lack of necessary comparison UV data from a different epoch
prevent a clear indication of a change.

\begin{figure}
\begin{center}
\epsscale{0.9}
\rotatebox{90}{
\plotone{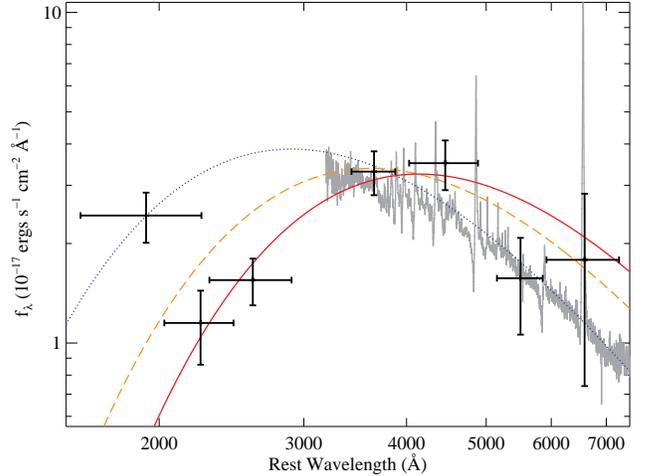}}
\caption{UV/Optical photometry of \ip\ during the fading event
immediately after maximum brightness.  The blue dotted, orange dashed,
and red solid curves correspond to 10,000, 8000, and 7000~K blackbody
spectra, respectively.  The 23~day spectrum is also plotted to show
the consistency with both the photometry and the 10,000~K blackbody.
All photometry is consistent with the 8000 and 10,000~K blackbody
spectra.  Ignoring the bluest (UVW2) filter, the data are also
consistent with the 7000~K blackbody.}\label{f:ipbb}
\end{center}
\end{figure}

Using the 15~day optical spectrum and 22~day NIR spectrum, we are able
to examine the SED of \ugc\ over nearly a decade in wavelength.
Between these dates, the light curve of \ugc\ was essentially
constant, having the same magnitude (within 1$\sigma$)
\citepalias{Smith09:outburst}.  Using the long wavelengths of the NIR
spectrum, our data are sensitive to any low-temperature thermal
components.

We fit a single blackbody to these data, ignoring regions with strong
line features and simultaneously fitting the scaling between the
optical and NIR spectra.  Doing this results in a best-fit temperature
of 6800~K.  This single blackbody consistently under-predicts the flux
at NIR wavelengths.  As a result, we have also attempted to fix the
spectrum with a double blackbody model.  This model, which produces a
much better fit, results with temperatures $T_{1} = 6950$~K and $T_{2}
= 2100$~K.  The full spectrum and associated fits are shown in
Figure~\ref{f:ubb}.

\begin{figure}
\begin{center}
\epsscale{0.9}
\rotatebox{90}{
\plotone{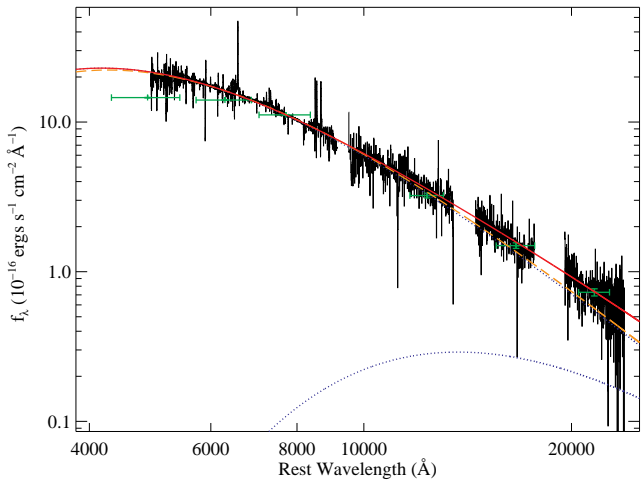}}
\caption{Optical/NIR spectrum of \ugc.  Single (6800~K; dashed orange
line) and double (2100 and 6900~K; solid red line) blackbody fits to
the spectrum are overplotted.  The individual components of the double
blackbody fit are shown as blue dotted lines.  The double blackbody is
a better fit to the data than the single blackbody.  Green points show
our photometry, which also shows a NIR excess.  The $g$-band flux is
below that of either curve, but that is likely the result of line
blanketing.}\label{f:ubb}
\end{center}
\end{figure}

\citetalias{Smith09:outburst} noted that \ugc\ had a (photometric) NIR
excess, but could not distinguish between circumstellar extinction and
dust emission.  To test the former case, we attempted to fit the
spectrum with a single blackbody, but with an additional extinction
term.  With $R_{V}$ fixed to 3.1, this model did not fit the data
well.  The model was able to sufficiently reproduce the data if we
allowed $R_{V} < 1$, which is unphysical.  We therefore conclude that
the NIR excess is likely due to dust emission.

Scaling our spectrum to our broad-band photometry, we can calibrate
the blackbody flux, which in turn constrains the ratio $R/D$, where
$R$ is the radius of the blackbody radiation and $D$ is the distance
to the object.  Using $D = 6 \pm 0.5$~Mpc, we find that the hot and
cool blackbodies have radii of $(1.50 \pm 0.16) \times 10^{14}$~cm and
$(4.3 \pm 0.4) \times 10^{14}$~cm ($13.0 \pm 1.1$~AU and $29 \pm
2$~AU), respectively.  The size of the cool emitting region is of the
same order of magnitude of the size of the Homunculus nebula
surrounding $\eta$~Car.

Following the prescription outlined by \citet*{Smith08:06jc} (and
references therein), we can measure the mass of the emitting dust.
Specifically,
\begin{equation}
M_{d} = \frac{400 \pi \rho R_{d}^{2}}{3 T_{d}^{2}},
\end{equation}
where $M_{d}$ is the dust mass, $R_{d}$ is the radius of the dust,
$T_{d}$ is the dust temperature, and $\rho$ is the dust density.  For
the values obtained from the spectra, $T_{d} = 2100$~K and $R_{d} =
4.3 \times 10^{14}$~cm, and assuming a dust grain density of $\rho =
2.25$~g~cm$^{-3}$, we find a dust mass of $M_{d} \approx 2 \times
10^{-8} M_{\sun}$.  Since there could be a significant amount of dust
emission at lower temperatures, this is a lower limit on the total
dust mass; however, it is worth noting that this measurement is orders
of magnitude less than the dust created in some SNe \citep[e.g.,][and
references therein]{Kotak09}.  We note depending on the dust
composition, the dust temperature may differ from the blackbody
temperature by hundreds of degrees.

The dust is very close to the star and its temperature is near the
limit of grain survival.  Given these conditions, it is very likely
that pre-existing circumstellar dust was heated and is emitting as it
is being vaporized, rather than newly formed dust emitting as it
cools.

%%%%%%%%%%%%%%%%%%
%%  Discussion  %%
%%%%%%%%%%%%%%%%%%

\section{Discussion}\label{s:disc}

\subsection{Different Massive Star Outbursts}

\ugc\ and \ip\ provide excellent examples of the diversity of massive
star outbursts.  \ugc\ increased its optical brightness by \about
1~mag during outburst and has a cool spectrum with many narrow
absorption lines and [\ion{Ca}{2}] emission.  It occurred near a star
cluster containing stars with initial masses of \about $25 M_{\sun}$
and shows evidence for a very cool ($T \approx 2100$~K) thermal
component that is radiated by circumstellar dust.  \ip\ had a more
massive and relatively isolated progenitor.  At peak, it had risen at
least 4~mag over the previous year, and its spectrum was hot and
dominated by H Balmer emission lines.  After having a significant
fading and rebrightening over three weeks, it developed high-velocity
absorption lines.  Its SED is consistent with no dust emission.

The high temperature and large increase in optical luminosity for \ip\
indicates that it was a true giant eruption akin to the 1843 eruption
of $\eta$~Car.  \ugc, on the other hand, has spectral characteristics
similar to that of S~Dor at maximum.  The relatively small increase in
optical luminosity may indicate that \ugc\ was the result of normal
S~Dor variability, but that it was a particularly luminous maximum.
While the largest normal variation of S~Dor stars vary by \about 3~mag
in the optical \citep{vanGenderen01}, \ugc\ has varied by at least
\about 5~mag (including the current eruption) over the last ten years
(Section~\ref{s:obs}; see also \citetalias{Smith09:outburst}).

The outbursts of \ngc\ and SN~2008S both had relatively low
temperature SEDs \citep{Berger09:ngc}, but neither had the forest of
absorption lines found in \ugc.  All three objects had [\ion{Ca}{2}]
emission, but as suggested by \citetalias{Smith09:outburst}, this may
be linked to the circumstellar environment, and particularly dust
destruction, rather than the event.  Our observations have shown that
there is dust in the circumstellar environment of the progenitor, and
that it was likely pre-existing dust that is in the process of being
vaporized.  Additionally, SN~1999bw had [\ion{Ca}{2}] emission
\citep{Garnavich99} and had an IR excess consistent with dust emission
at late times \citep{Sugerman04}.  All four massive star outbursts
with observed [\ion{Ca}{2}] emission (SN~1999bw, SN~2008S, \ngc, and
\ugc) have evidence of circumstellar dust.  We do note that SN~2000ch
had a \about 7000~K spectrum and an infrared excess consistent with
dust emission, but no strong [\ion{Ca}{2}] emission was detected in
the relatively low signal-to-noise spectra presented by
\citet{Wagner04}.  It is therefore possible to have a cool object and
circumstellar dust yet not have [\ion{Ca}{2}] emission.

Although \ugc, \ngc, and SN~2008S have circumstellar dust and similar
temperatures, other than the narrow H Balmer and [\ion{Ca}{2}] (which
is linked to the presence of circumstellar dust) emission, the spectra
and progenitors are not particularly similar.  Particularly, \ngc\ and
SN~2008S had relatively featureless spectra and progenitors with
initial masses of 10 -- $25 M_{\sun}$, while \ugc\ had spectrum
dominated by narrow lines and a more massive progenitor ($\gtrsim 25
M_{\sun}$).  Additional data are necessary to determine if the
outburst mechanisms in these objects are similar.

Using \ip\ and \ugc\ as examples, there appears to be two distinct
elements that determine the observational properties of massive star
outbursts.  The first is the temperature of the outburst, which may be
related to the increase in luminosity, the instability that causes the
eruption, the width of the emission lines, and possibly the energetics
of the outburst and if there is an explosion (see
Section~\ref{ss:exp}).  This directly determines the shape of the
optical SED, the ionization, if there is a forest of absorption lines,
and possibly the shape of the line profiles, if there is high-velocity
absorption.  The other characteristic is the amount of circumstellar
dust, which may cause strong Ca emission (and particularly
[\ion{Ca}{2}] emission) and will determine the shape of the SED at
longer wavelengths.  \ugc\ and \ip\ would occupy very different
regions of the parameter space created by these two dimensions.  \ngc\
and SN~2008S would be close to \ugc, while LBV giant eruptions such as
SN~1997bs would be close to \ip.

It remains to be seen if there are hot massive star outbursts with a
large amount of circumstellar dust or if there are cool massive star
outbursts with little circumstellar dust.  $\eta$~Car has $0.125
M_{\sun}$ of dust surrounding it \citep{Smith03}; if it were to have
another giant eruption today, would it be cool?  SN~1999bw had
[\ion{Ca}{2}] emission and displayed dust emission at late times; was
it hot?  An IR survey of recent massive star outbursts with good
spectroscopic coverage may provide these answers.  In the future,
optical and NIR observations may be sufficient to determine these
characteristics for other massive star outbursts.

\subsection{\ip: A Supersonic Explosion}\label{ss:exp}

The spectra of \ip\ have absorption attributed to high-velocity (up to
\about 7000~\kms) material (see Section~\ref{sss:halpha}).  Contrary
to what is expected from a single outburst or explosion, the velocity
of the absorption feature increases with time.  In a typical SN, the
ejecta naturally follow a Hubble law with the highest velocity
material being the most distant from the explosion site.  Spectral
lines have a blueshifted absorption due to the scattering processes in
the photosphere of the SN.  Low velocity material is hidden behind the
photosphere, only to be revealed at later times.  As the photosphere
recedes, the highest-velocity material becomes optically thin,
resulting in the blueshifted velocity of a spectral line to decrease
with time.  Since the absorbing material must be at just slightly
larger radii than the photosphere, the high-velocity material must
have been ejected during the eruption.  (If the absorbing material
were from a previous eruption, the ejecta from the more recent
eruption would have had to be moving even faster.)

It is possible that the high-velocity absorption is a component of
P-Cygni features from the ejected material.  The Lorentzian profile
slightly underestimates the emission flux in the 86~day spectrum (see
Figure~\ref{f:halpha}), which may be the result of P-Cygni emission
contributing to the line.  Since the high-velocity absorption is
coming from the ejecta, the outburst of \ip\ must have been extremely
energetic, expelling a large amount of material at very high
velocities.  However, for the velocity to {\it increase} with time,
either the ejecta must not follow a Hubble expansion or the radius of
the photosphere (in velocity space) must somehow increase with time.

In a single explosion, the ejecta naturally follow a Hubble law;
however, multiple explosions can change the velocity profile of the
ejecta.  If two explosions occurred in short succession, one can
produce the inverted velocity gradient seen in \ip.  In this toy
model, the photosphere would recede into the ejecta of the first
explosion, but at some point the fastest-moving ejecta of the second
explosion would overtake the photosphere, increasing the velocity.  If
there are no other explosions, the velocity of the absorption would
decrease from there.

The photometric behavior of \ip\ is consistent with this picture.  The
first explosion would produce the fast rise to maximum.  As noted by
\citetalias{Smith09:outburst}, the timescale of the fading is much
shorter than the timescales for many physical processes such as dust
extinction.\footnote{The calculation by \citepalias{Smith09:outburst}
for the time until dust formation for \ip\ assumes a velocity of
500~\kms.  Although the ejecta are moving much faster than this
assumed value, they would need to have a velocity of $\gtrsim
20,000$~\kms\ to reach the sublimation radius at the time of fading.}
This behavior is very similar to that of SN~2000ch, which brightened
by 2.1~mag in 9~days to maximum, then immediately faded by 3.4~mag in
7~days, immediately followed by a 2.2~mag rise in 4~days, after which
the magnitude stayed relatively constant \citep{Wagner04}.  Spectra of
SN~2000ch taken during the fading and on its plateau show no strong
evidence for high-velocity ejecta, but the spectra may not be of high
enough quality to see these features.

\citetalias{Smith09:outburst} hypothesized that rapid fading may
have been caused by an optically thick shell being ejected after the
first outburst.  If this process did occur in \ip, then there are
several implications: (1) the velocity of the absorption should
eventually decrease, (2) the interaction of the ejecta from the two
explosions could be a significant source of X-ray and radio emission,
and (3) the X-rays might excite certain elements producing
high-excitation lines such as \ion{He}{2} in optical spectra.  Our
X-ray limit of $L_{X} < 4.8 \times 10^{38}$~ergs~s$^{-1}$ taken during
the minimum is not particularly constraining.  We do not detect any
\ion{He}{2} $\lambda 4686$ emission in the 4 or 24~day spectra;
however, there is a low significance detection of a line consistent
with \ion{He}{2} $\lambda 4686$ emission in the 86~day spectrum.
Additional spectroscopy is necessary to determine the late-time
velocity gradient.

\section{Conclusions}\label{s:con}

We have presented extensive UV, optical, and NIR data for two
transients, \ip\ and \ugc.  Although these events appear to be similar
phenomena (luminous outbursts of massive stars), the details of the
events show that there are many differences.  These differences
provide examples of the diversity of such events.

A previous study of these events, \citetalias{Smith09:outburst},
provided an initial analysis of the object.  Although the two studies
agree on many points, our interpretation of the entire data set is
somewhat different than that of \citetalias{Smith09:outburst}.  In
particular, we agree that based on pre-outburst {\it HST} imaging,
historical light curves, and outburst spectroscopy, the progenitors of
\ip\ and \ugc\ are LBVs with masses of $\gtrsim 60$ and $\gtrsim 25
M_{\sun}$, respectively.  We also agree that the spectra of the two
events are significantly different, but consistent with known LBVs or
LBV outbursts.  While \ugc\ had a cooler spectrum with a forest of
absorption lines reminiscent of a F-type supergiant (similar to S~Dor
in its high state), \ip\ had a hot spectrum and exhibited mainly H
Balmer emission (similar to other LBV giant eruptions).  The spectral
characteristics (particularly [\ion{Ca}{2}] emission) and
circumstellar dust link \ugc\ to the lower-mass, dust-obscured
progenitors of \ngc\ and SN~2008S.  We agree that the progenitors of
these objects are all massive stars and may have many characteristics
similar to those of the LBV class, which could extend the mass range
for LBV-like activity to relatively low-mass stars.

However, there are distinct differences between the analyses of
\citetalias{Smith09:outburst} and of that presented here.
Specifically, the initial mass ranges for the progenitors are slightly
different in the two studies, with \citetalias{Smith09:outburst}
estimating 50 -- $80 M_{\sun}$ (instead of $\gtrsim 60 M_{\sun}$) and
$\gtrsim 20 M_{\sun}$ (instead of $\gtrsim 25 M_{\sun}$) for \ip\ and
\ugc, respectively.  The differences lie in the conversion from \hst\
filters to Bessell filters and the adapted color range for the
progenitor of \ip, and the additional information provided by stars in
the vicinity of the progenitor of \ugc.

Our interpretation of the exact nature of \ugc\ differs from that of
\citetalias{Smith09:outburst}.  While \citetalias{Smith09:outburst}
contends that this object is a true giant eruption of an LBV, we
question this assertion and propose that it may be the result of
extreme S~Dor variability.

While \citetalias{Smith09:outburst} found an NIR excess for \ugc,
hypothesizing that there may be dust emission as it is vaporized, we
find more conclusive evidence for this scenario through our NIR
spectroscopy.  The NIR spectrum is consistent with an additional
blackbody with $T \approx 2100$~K, but is inconsistent with reddening
by dust.  The presence of this dust indicates that the initial mass
estimate of the progenitor (based on optical {\it HST} imaging) is a
lower limit, and that the true initial mass is likely much larger.

In addition to what was discussed by \citetalias{Smith09:outburst}, we
have also detected high-velocity absorption in the spectra of \ip,
indicative of an explosion (as opposed to subsonic outburst).  The
absorption has an inverse velocity gradient suggesting multiple
explosions in short succession.  The rapid fading and brightening
shortly after maximum brightness noted by
\citepalias{Smith09:outburst} is consistent with multiple explosions,
where a second explosion ejects an optically thick shell that
temporarily dims the object.

We also note the spectroscopic similarity of \ugc\ and SN~1994W, which
\citet{Dessart09} has previously suggested was not a true SN that
destroyed its progenitor star.

\ip\ and \ugc\ are very different manifestations of a similar
phenomenon: extreme brightening of massive stars.  With these objects
and similar events (such as $\eta$~Car, \ngc, and SNe~1961V, 1954J,
1997bs, 2000ch, 2002kg, and 2008S), we show that luminous outbursts of
massive stars are very heterogeneous.  Some of this diversity is
likely linked to the instability that causes the eruption, while some
is caused by the circumstellar environment.  Additional observations
of new massive star eruptions are necessary to determine the physical
mechanisms of the eruptions, the content of the circumstellar
environments, and whether the two are physically connected.

\begin{acknowledgments} 

{\it Facilities:} 
\facility{ARC (TripleSpec), FMO:31in (FanCam), Gemini:Gillett (GMOS),
HST (WFPC2), Magellan:Baade (IMACS), Magellan:Clay (MagE), MMT (Blue
Channel), Swift (UVOT, XRT)}

\bigskip
R.J.F.\ is supported by a Clay Fellowship.  O.D.F.\ is grateful for
support from NASA GSRP, ARCS, and VSGC.  E.M.L.\ is supported in part
by a Ford Foundation Predoctoral Fellowship.

We are indebted to the staffs at the APO, Gemini, Magellan, and MMT
Observatories for their dedicated services.  We thank K.\ Olsen and
R.\ McDermid for obtaining some of the data presented in the paper.
We thank R.\ Chornock and R.\ Kirshner for stimulating discussions
about the transients.

Based on observations made with the NASA/ESA Hubble Space Telescope,
obtained from the data archive at the Space Telescope Science
Institute.  STScI is operated by the Association of Universities for
Research in Astronomy, Inc.\ under NASA contract NAS 5-26555.  Based
in part on observations obtained at the Gemini Observatory, which is
operated by the Association of Universities for Research in Astronomy,
Inc., under a cooperative agreement with the US National Science
Foundation on behalf of the Gemini partnership: the NSF (United
States), the Science and Technology Facilities Council (United
Kingdom), the National Research Council (Canada), CONICYT (Chile), the
Australian Research Council (Australia), Minist\'{e}rio da Ci\^{e}ncia
e Tecnologia (Brazil) and Ministerio de Ciencia, Tecnolog\'{i}a e
Innovaci\'{o}n Productiva (Argentina); the 6.5 meter Magellan
Telescopes located at Las Campanas Observatory, Chile; the MMT
Observatory, a joint facility of the Smithsonian Institution and the
University of Arizona; the Fan Mountain Observatory 0.8 meter
telescope; and the Apache Point Observatory 3.5 meter telescope, which
is owned and operated by the Astrophysical Research Consortium.  We
acknowledge the use of public data from the {\it Swift} data archive.

\end{acknowledgments}

\bibliographystyle{fapj}
\bibliography{astro_refs}

%\eject

\end{document}